\documentclass[usenatbib]{mn2e}

\newcommand{\be}{\begin{eqnarray}}
\newcommand{\bee}{\begin{enumerate}}
\newcommand{\bit}{\begin{itemize}}

\def\bkt#1{\left(#1\right)} % normal brackets
\def\bkts#1{\left[#1\right]} % square brackets
\def\bkta#1{\langle#1\rangle} % angle brackets

\def\ee{\end{eqnarray}}
\newcommand{\eee}{\end{enumerate}}
\def\eg{\textit{e.g.} }
\newcommand{\eit}{\end{itemize}}
\def\fnl{f_{\mbox{\scriptsize NL}}}
\def\ff{\phantom{.}}

\newcommand{\ii}{\textit}

\def\lab{\label}

\newcommand{\mb}{\mathbf}
\def\mc#1{\mathcal{#1}}
\newcommand{\mmm}{\medskip}

\newcommand{\no}{\noindent}
\newcommand{\nn}{\nonumber}

\def\pr{\prime}
\def\re#1{(\ref{#1})}

\newcommand{\sss}{\smallskip}
\def\sub#1{_{\mbox{\scriptsize{#1}}}}
\def\sun{\odot}

\usepackage{amsmath}
\usepackage{amsfonts}
\usepackage{amssymb}
\usepackage{graphicx}
\usepackage{epsfig}
\usepackage{hyperref}

\usepackage{bm}

\begin{document}

\title[21cm radiation from minihalos as a probe of small non-Gaussianity]{The 21cm radiation from minihalos as a probe of small primordial non-Gaussianity}
%\title[Detecting non-Gaussianity with 21cm signal from minihalos]{Detecting small primordial non-Gaussianity with 21cm signal from minihalos}
\author[Chongchitnan and Silk]{Sirichai Chongchitnan$^{1,2},$	Joseph Silk$^{1,3,4}$\\
	$^1$ Oxford Astrophysics, Denys Wilkinson Building, Keble Road, Oxford, OX1 3RH,\\
	$^2$ School of Engineering, Computing and Applied Mathematics, University of Abertay Dundee, Bell St., Dundee, DD1 1HG,\\
         $^3$Institut d'Astrophysique, UPMC, 98bis Boulebard Arago, Paris 75014,\\
         $^4$Department of Physics and Astronomy, The Johns Hopkins University, 3701 San Martin Drive, Baltimore MD 21218.}

\date{May 2012}
\pagerange{\pageref{firstpage}--\pageref{lastpage}}
\pubyear{2012}

\maketitle

\begin{abstract}
%%%
We present a new probe of primordial non-Gaussianity via the 21cm radiation from  minihalos at high redshifts. We calculate the fluctuations in the brightness temperature (measured against the cosmic microwave background) of the 21cm background from minihalos containing HI at redshift $\sim6-20$, and find a significant enhancement due to small non-Gaussianity with amplitude $\fnl\lesssim1$. This enhancement can be attributed to the nonlinear bias which is strongly increased in the presence of non-Gaussianity. We show that our results are robust against changes in the assumed mass function and some physical aspects of minihalo formation, but are nevertheless sensitive to the presence of strong radiation sources within or around the minihalos. Our findings are relevant for constraining and searching for small primordial non-Gaussianity with upcoming radio telescopes such as LOFAR and SKA.

%Minihalos are earliest virialised structures of neutral hydrogen clouds. The signals from the spin-flip transitions in these clouds may be detectable by the upcoming large arrays of radio telescopes such as the Square Kilometre Array.

%We investigate the effects of small primordial non-Gaussianity on the 21-cm signals from minihalos at redshift $\lesssim20$. For non-Gaussianity parametrized by $\fnl$, we show that $\fnl=\mc{O}(1)$ gives rise to a significant enhancement in the 21-cm radiation seen in emission against the cosmic microwave background. The enhanced  fluctuations in this signal across the sky may be detectable by the upcoming radio surveys such as the Square-Kilometre Array. 

%%%
\end{abstract}

\section{Introduction}

In this era of ``precision cosmology'', many open questions in cosmology will be addressed with the highly anticipated results from the \ii{Planck} satellite\footnote{\texttt{www.rssd.esa.int/planck}} and a host of other ambitious ground and space-based experiments. One of the key goals of these experiments is to establish the statistical nature of the primordial cosmological perturbations, which, according to the single-field, slow-roll model of inflation, should be very close to a Gaussian field (see \eg \cite{maldacena}, for reviews see \cite{bartolo,chen}). A detection of any primordial non-Gaussianity (NG) would therefore signify a deviation from the simplest model of inflation and hint at new physics in the early Universe.

At present, the most stringent constraint on NG comes from measurements of the anisotropies in the cosmic microwave background (CMB). In the ``local" model of NG parametrized by a constant amplitude, $\fnl$, the \ii{WMAP} satellite reported a limit $\fnl=32\pm42$ ($2\sigma$) \citep{komatsu} whilst \ii{Planck} is expected to improve this limit to $|\fnl|\lesssim5$.

On smaller physical scales, galaxy clusters have been shown to be effective in constraining NG via the number counts of rare objects and measurements of the bias (see \eg \cite{scoccimarro,dalal,slosar,desjacques, mebias}). A positive $\fnl$, for instance, will increase the number of galaxy clusters at high redshifts and induce scale-dependence in the bias. Realistically, however, uncertainties in the mass-observable relation, redshift determination and other systematics will most likely limit the constraining power of cluster surveys to 
$|\fnl|>10$. \cite{slosar}, for example, obtained the $2\sigma$ limit of $-29<\fnl<70$ from the analysis of various large-scale-structure datasets.

%using the rarest, most massive clusters of typically $\sim10^{15}M_\sun$ \citep{cayon,me}.

More recently, it has been shown that NG leaves imprints in the 21cm radiation due to spin-flip transitions of neutral hydrogen during the epoch of reionization. This transition occurs at an excitation temperature $T_*=68$ mK and rest-frame frequency $\nu_0=1.42$ GHz (now redshifted to the radio band). These 21cm signals are expected to be measured over a wide range of redshifts by radio facilities such as LOFAR\footnote{Low-Frequency Array (\texttt{www.lofar.org})}, MWA\footnote{Murchison Widefield Array (\texttt{www.mwatelescope.org})} and SKA\footnote{Square-Kilometre Array (\texttt{www.skatelescope.org})}. It was first shown in \cite{cooray} that an ideal measurement of the 3-point correlation in the 21cm fluctuations from $z\sim50$ can, in principle, probe $\fnl$ as small as $10^{-2}$ (see also \cite{pillepich}). \cite{joudaki} considered a more conservative approach of measuring the 21cm power spectrum and concluded that $|\fnl|\sim10$ is within reach of the next generation of radio telescopes. \cite{tashiro1} analysed the number counts of ionized bubbles in 21cm maps and found them to be sensitive to $|\fnl|\sim100$. Most recently, \cite{tashiro} argued that the correlation between the CMB and 21cm fluctuations will be capable of detecting $|\fnl|\sim100$ and will be useful for removing foregrounds and systematics.

% On MH
In this work, we present a new connection between NG and 21cm radiation, namely, via the 21cm signals from minihalos (MHs), which are some of the earliest cosmic  structures to have formed post-recombination. MHs are typically virialised clouds with mass $10^4-10^8 M_\sun$, comprising dark and baryonic matter,  which have cooled to below $\sim 10^4$ K. These MHs host such a high density of neutral hydrogen that their 21cm signal can appear in emission or absorption against the CMB, resulting in a 21cm ``forest" of spectral lines \citep{iliev,shapiro,furlanetto,meiksin}. Through the changes in number density and bias, we will show that the fluctuations in the 21cm emission from MHs provides a new probe of small NG with $\fnl\lesssim1$ and is potentially detectable by LOFAR and SKA.

% Cosmo params
Throughout this work, we assume a flat $\Lambda$CDM cosmology with matter density $\Omega_m=0.276$, baryon density $\Omega_b=0.046$, Hubble constant $h=0.7$, spectral index $n_s = 0.96$ and power-spectrum normalisation $\sigma_8=0.81$.

%tomographic redshift slices between the last scaterring surface and $z\sim1$ typical of massive clusters.

\section{21cm emission from minihalos}

We follow the basic treatment of halos in \cite{shapiro1,iliev1} where a MH of a given mass is modelled as a ``truncated isothermal sphere" with radius $r_t$, temperature $T_K$, dark-matter velocity dispersion $\sigma_V$, and density profile $\rho(r)$.

% Furlanetto and Loeb (0312435) : 21cm emission from shocked gas in the IGM can overwhelm MH emission... tested against simulation by Shapiro etal who found that the former is dominant.

We begin by considering a single MH. Its 21cm signal may appear in emission or absorption against the CMB depending on the spin temperature, $T_S$, which is determined by 1) energy exchanges between HI-bound electrons and CMB photons, 2) collisions between atoms, 3) interactions between electrons and Lyman-$\alpha$ (Ly$\alpha$) photons through the Wouthuysen-Field effect \citep{wouthuysen,field}. We can write $T_S = (T\sub{CMB} +y_\alpha T_\alpha +y_c T_K)/(1 + y_\alpha+y_c),$ where $y_\alpha, y_c$ are the Ly$\alpha$ and collisional coupling constants \citep{madau,allison} and $T_\alpha\approx T_K$ is the Ly$\alpha$ colour temperature. Assuming for now that bright UV and X-ray sources have yet to form, or that the MH is isolated from such sources, we can set $y_\alpha=0$.
 
%\be y_c = {T_* n_{HI}\kappa(1-0)\over T_K A_{10}},\ee % nH or nHI??
%where the coefficient $\kappa(1-0)$ (as a function of $T_K$) is tabulated in Allison and Dalgarno ApJ 158 (1969).

The observed brightness temperature at comoving distance $r$ from the centre of the halo is given by
\be T_b(r)= T\sub{CMB}(0)e^{-\tau (r)}+\int_0^{\tau}T_S e^{-\tau^\pr}d\tau^\pr,\ee
where $T\sub{CMB}(z)=2.73(1+z)$ K. The optical depth, $\tau$, of neutral hydrogen to photons at rest-frame frequency, $\nu$, along a line of sight with impact parameter, $\alpha$ (in unit of $r_t$,) from the centre of the MH can be expressed as \citep{furlanetto}
\be
% {\partial \tau \over \partial \ell} &=& {3c^2 A_{10}T_*\over 32\pi   \nu^2} \cdot {n\sub{HI}(\ell)\phi(\nu,\ell)\over T_S(\ell)},\lab{tau}\\
\tau (\nu)&=& {3c^2 A_{10}T_*\over 32\pi\nu_0^2  }\int_{-\infty}^{\infty} {n\sub{HI}(\ell)\phi(\nu,\ell)\over T_S(\ell)} dR,\lab{tau}\ee
where $A_{10}=2.85\times10^{-15}$ s$^{-1}$ and $R$ and $\ell$ are radial comoving distances satisfying $\ell^2=R^2+(\alpha r_t)^2$, with $R=0$ at the centre of the MH. The number density of neutral hydrogen in the MH is $n\sub{HI}\approx (1-Y)(\Omega_b /\Omega_m)(\rho/m_H)$, where $Y$ is the helium fraction and $m_H$ is the mass of a hydrogen atom. 
%$Y\approx0.24$.
%The number density of neutral hydrogen can be expressed as $n\sub{HI}=x\sub{HI}(z)\times n_H(z)$ where $x\sub{HI}$ is the neutral fraction of hydrogen. As a first approximation we set $x\sub{HI}=1$.
%(and maybe use a sensible model with , $x=1$ when z drops below 6?). 
%In the IGM, we can write \be n_H(z)= 1.9\times 10^{-7} \mbox{cm}^{-3}(1+z)^3.\ee In the MH, 
The intrinsic line profile $\phi(\nu)$ is modelled as a Doppler-broadened form 
$\phi(\nu)=(\Delta \nu \sqrt{\pi})^{-1}\exp\bkt{-[(\nu-\nu_0)/\Delta \nu]^2}$, with $\Delta\nu = (\nu_0/c)\sqrt{2k_B T_K(z)/m_H},$ and
 $k_B$ the Boltzmann constant.

%and $v\sub{LOS}$ is the peculiar velocity of the MH projected along the line of sight. No redshifting between the observer and the halo has yet been taken into account.

In the special case when the line is unbroadened, $\phi(\nu)=\delta(\nu-\nu_0)$, 
%we can convert \re{tau} to an integral in $z$ using \re{elltoz}, and 
the optical depth reduces to that of the unperturbed IGM patch at redshift $z$ \citep{madau}
\be \tau\sub{IGM}(z)= {3c^3 A_{10}T_* n\sub{HI}(z)\over 32\pi\nu_0^3 T_S(z)H(z)}.\lab{tauigm}\ee 
%Finally, how does $T_K$ vary with $\nu$ and $z$. In terms of the radius $r$,
%\be T_K(r)=\begin{cases}
%T_K\super{TIS}, \ff r\leq r_t,\\
%T_K\super{IGM}, \ff r>r_t,
%\end{cases}
%\ee
%where % eq 81 of iliev and shapiro
%\be T_K\super{TIS}&=& 7.843\times 10^5 \bkt{\mu\over 0.59}\bkt{M\over 10^{12}h^{-1}M_\sun}^{2/3}F(z\sub{MH}) \ff \mbox{K},\\
%T_K\super{IGM}&=& 
%\begin{cases}
%T\sub{CMB}(z)\ff , z>500,\\
%1.46(1+z)^{1.1} \ff \mbox{K}, \ff 300<z\leq500,\\
%0.146(1+z)^{1.5} \ff \mbox{K}, \ff 60<z\leq300,\\
%0.023(1+z)^{1.95} \ff \mbox{K}, \ff 6<z\leq60,\\
%\end{cases}\ee % adiabatic cooling.. eq 12 of shapiro 0512516
%The mean molecular weight $\mu=1.22$ for a neutral hydrogen atom.
Using \re{tau}-\re{tauigm}, we can rewrite the brightness temperature as 
\be T_b(\nu)&=& T\sub{CMB}e^{-\tau (\nu)}+\int_{-\infty}^\infty T_S(\ell) e^{-\tau(\nu,R)}{\partial\tau\over \partial R}dR,\\
\tau(\nu,R)&=& \tau\sub{IGM}+{3c^2 A_{10}T_*\over 32\pi\nu_0^2  }\int_{-\infty}^{R} {n\sub{HI}(\ell^\pr)\phi(\nu,\ell^\pr)\over T_S(\ell^\pr)} dR^\pr.\ee
%where $(\ell^\pr)^2=(R^\pr)^2+(\alpha r_t)^2.$ 
Finally, the observed 21cm brightness temperature of a single MH with respect to the CMB is given by
\be \delta T_b = {\bkta{T_b}  \over 1+z}- T\sub{CMB}(0),\lab{tbhalo}\ee
where $T_b$ is averaged over the halo cross-section $A=\pi r_t^2$.

%Let $A=\pi r_t^2$ denote the cross section of each halo. The face average temperature of each halo is then $\bkta{T_b}=A^{-1}\iint T_b(r) dA = 2\int_0^{1} \alpha T_b(\alpha r_t) d\alpha.$

%Note that the integral in $r$  may be evaluated by applying Leibniz's theorem on \re{rnu} so that 
%\be {dr\over d\nu}= -{c\nu_0\over \nu^2 H(\nu/\nu_0)}\ee

%\subsection{Emission from a population of minihalos}

If we now consider an ensemble of MHs in the mass range $[M\sub{min},M\sub{max}]$, the mean 21cm  emission from an ensemble is given by \citep{iliev}
\be \overline{\delta T_b}= {c(1+z)^4\over \nu_0 H(z)}\int_{M\sub{min}}^{M\sub{max}}\Delta \nu\sub{eff} \ff \delta T_{b}(M) A {dn\over dM} dM.\ee
where $\Delta \nu\sub{eff}=[\phi(\nu_0)(1+z)]^{-1}$ is the effective redshifted linewidth. Various prescriptions for the mass function, $dn/dM$, will be compared later on. We set $M\sub{max}$ to correspond to the virial temperature of $10^4$ K  whilst $M\sub{min}$ is set by the Jeans mass, $M_J$. 
%Uncertainties in these thresholds will also be explored in section \ref{later}.

%\subsection{Anisotropy}
%The survey volume which can be modelled as a pencil beam of radius $R$ and length $L$. For a survey with frequency bandwidth $\Delta\nu$ and angular size $\Delta\theta$, we can write $L\approx c(1+z)H(z)^{-1}(\Delta \nu/\nu)\sub{obs}$ and $R\approx r(z)\Delta\theta /2$ (where $r(z)$ is the comoving distance corresponding to redshift $z$) [cite Tozzi 9903139]. The variance, $\sigma_p^2$, of the density field the beam can be expressed as\footnote{The Fourier transform convention follows that of Valageas}
%\be \sigma_p^2 = 4\pi\int_0^1dx \int_0^\infty {dk\over k} \ff \bkts{j_0(kLx/2)W_\perp(kR\sqrt{1-x^2})(1+\Omega_m(z)^{0.6}x^2)}^2\mc{P}(k,z)\ee
%where $\mc{P}(k)$ is the dimensionless power spectrum evaluated at redshift $z$ and $W_\perp(y)=2J_1(y)/ y$ is the transverse filter function for a disk radius $R$ [cite Kaiser and Peacock apj 379, 482 1991]

The key observable relevant for the upcoming radio arrays is the \ii{rms} fluctuations in the 21cm emission. 
For a pencil-beam survey with frequency bandwidth $\Delta\nu$ and angular size $\Delta\theta$, the amplitude of the $3\sigma$ fluctuation is
\be \bkta{\delta T_b^2}^{1/2}=3\sigma_p(\Delta\nu,\Delta\theta)\beta(z)\overline{\delta T_b}\lab{fluct},\ee (see e.g. \cite{dodelsonbook} for the calculation of the variance $\sigma_p$ in a cylinder). Here, $\beta(z)$ is the weighted average of the bias $b(M,z)$, defined as the ratio of the 2-point correlation for density peaks corresponding to a MH of mass $M$, and that of dark-matter density fluctuations (detail in the next section). 
\be \beta(z)= {\int_{M\sub{min}}^{M\sub{max}} b(M,z) \mc{F}(m){dn\over dM} dM \over \int_{M\sub{min}}^{M\sub{max}}\mc{F}(m){ dn\over dM} dM}.\lab{weight}\ee where $\mc{F}(m)\propto T_b r_t^2\sigma_V$ is the effective flux from the MHs. %[see \cite{iliev} for detail]. 

%, weighted by the received line-integrated flux $\mc{F}$, where 
%\be \mc{F}={2\pi k_B\nu_0^2\over c^2D_A^2(1+z)^3} r_t^2T_b(\nu_0) \Delta\nu\sub{eff}.\ee
%The flux received from each MH in the range $[M,M+dM]$, integrated over the bandwidth, is proportional to the face-average brightness temperature $\bkta{T_b}$ and also the effective line width $\Delta \nu\sub{eff}$. Let $\mc{F}(m)=\bkta{T_b}\sigma_V$. We can then define the weighted average of the bias as

%\be \beta(z)= {\int_{M\sub{min}}^{M\sub{max}} b(M,z) \mc{F}(m){dn\over dM} dM \over \int_{M\sub{min}}^{M\sub{max}}\mc{F}(m){ dn\over dM} dM}\ee, 

\section{Effects of $\fnl\lesssim1$}

If \ii{Planck} rules out $|\fnl|>$ a few, then it would appear extremely difficult for large-scale-structure probes to ever improve on, or even corroborate, $\fnl$ constraints from the CMB (unless $\fnl$ is $k$-dependent, see \eg \cite{loverde}). Unlike galaxy clusters, MHs are neither very rare nor very massive, so naively we expect the enhancement in their number counts from $\fnl\lesssim1$ to be undetectably small. However, at redshift $\gtrsim6$, these MHs are strongly biased  nonlinear objects and the scale-dependent effects on the bias can be much more dramatic than those on clusters.

In the Gaussian case, \cite{iliev2} used the nonlinear bias approach of \cite{scannapieco} (based on excursion set theory) to study 21cm emission from MHs. They found very good agreement for $\bkta{\delta T_b^2}$ between analytic prediction and simulation and concluded that, for all practical purposes, one could, for instance, use the bias obtained in \cite{mo} (using the peak-background split approach). In this work, we extend this line of investigation to non-Gaussian scenarios. 

A number of previous authors have investigated the effect of non-Gaussianity on the \ii{Fourier-space} bias, $b(k)$, defined as the ratio of the power spectrum for density peaks and that of dark matter (\eg \cite{dalal,matarrese,wagner}). However, an arguably more intuitive measure of the bias is in real space, where we can directly obtain information on the clustering amplitude of density peaks separated by comoving distance $r$. Following the pioneering work of \cite{kaiser}, we define
\be b^2(r) = {\xi\sub{pk}(r)\over\xi(r)},\lab{biasr}\ee
where $r$ is the comoving length in Eulerian space. The correlation function $\xi(\mb{x_1},\mb{x_2})=\bkta{\delta(\mb{x_1}),\delta(\mb{x_2})}$ where $\delta(\mb{x})$ is the overdensity field and $r= |\mb{x_1}-\mb{x_2}|$. Similarly, $\xi\sub{pk}$ is the 2-point correlation of density peaks corresponding to MHs of mass $M$. Whilst the real and Fourier space biases deal with the same physics of clustering in the overdensity field (and they are indeed equivalent in the Gaussian case), the real-space bias is related directly to the joint probability distribution of finding two overdense regions exceeding a collapse threshold in a given volume \citep{kaiser}. As NG, by definition, distorts this probability distribution from the Gaussian, the change in the real-space bias seems a natural, measurable quantity which can be calculated, for instance, using a saddle-point expansion \citep{valageasA} or Edgeworth expansion about the Gaussian \citep{mebias}. The Fourier bias, on the other hand, would be more useful when working with the power spectrum and bispectra from different NG shape templates \cite{matsubara}, or when redshift-space distortions are incorporated \cite{mao}. In this work, however, the \ii{real-space} bias suffices for the calculation of the 21cm fluctuations \re{fluct}.

To date there have only been a handful of calculations of the real-space bias in the presence of NG (and unfortunately $b(k)$ and $b(r)$ are not related via a straightforward Fourier transform). One such calculation is the saddle-point approach in \cite{valageasA,valageasB}. % whose approach involves a saddle point expansion of the bivariate probability distribution of the density contrast $P(\delta_1,\delta_2)$ in powers of $\fnl$. 
Whilst this formalism has been shown to agree with simulations of massive halos ($\sim10^{13}M_\sun$), it has yet to be tested against simulations of MH-scale resolution. We recognise that there may be limitations to the saddle-point formalism in the strongly non-linear regime. At the same time, there is not yet any convincing analytic model for the non-Gaussian bias in this regime, and thus we appeal to the saddle-point approach in this work as a first analytical approach to the problem. Recent progress in non-Markovian excursion-set theory \citep{adshead,musso,paranjape} should soon allow a more accurate calculation of the NG bias down to much smaller masses, and high-resolution simulations will be needed to elucidate the gas physics on such scales.

In \citep{me}, we studied the saddle-point approach in detail and found that when  $b(r)$ is averaged over all separation lengths $r$ within the volume that the MHs occupy (which in this case is a sphere with radius $L$), the result is the volume-averaged bias, $b(M,z)$, which is well-approximated by  
\be b(M,z)\approx[1+(6/5)\fnl K(z)L^2]b_G(M,z),\lab{kz}\ee where $b_G$ is the Gaussian bias and the constant $K$ can be determined from calculating $b(r)$ on some fixed scale. $K(z)$ roughly grows as a linear function of $z$ as shown in the top panel of Fig. \re{biases}.

\begin{figure}
\centering
\includegraphics[width= 11cm, angle = -90]{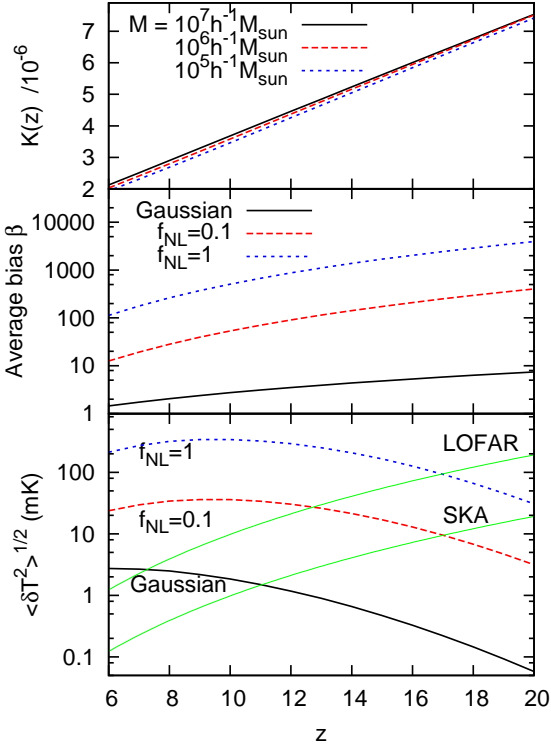}
\caption{\ii{Top:} The function $K(z)$ (see Eq. \ref{kz}) for a range of mass scales, $M$. $K(z)$ has a very weak dependence on $M$ and is independent of $\fnl$. \ii{Middle:} The volume-averaged bias, $\beta(z)$, of minihalos (Eq. \ref{weight}) with $\fnl=$ 0 (solid/black), 0.1 (long dashed/red) and 1 (short dashed/blue). \ii{Bottom:} The fluctuations in the 21cm brightness temperature, $\bkta{\delta T_b^2}^{1/2}$ (Eq. \ref{fluct}), assuming a bandwidth $\Delta\nu=1$MHz and angular diameter $\Delta\theta=9^\pr$ for the same range of $\fnl$. Overlaid (thin solid/green) are the noise thresholds for LOFAR and SKA-like experiments as discussed in the text.}
\label{biases}
\end{figure}

Using these results and combining it with the flux-weighting \re{weight}, we plot the average bias, $\beta(z)$, for $\fnl = 0,0.1$ and 1 in the middle panel of Fig. \ref{biases}. Clearly $\beta$ is sensitive to $|\fnl|\leq1$, which boosts the integrand in the nominator of \re{weight} particularly on mass scales around $M\sub{max}$. We have also checked that the ``Gaussian" curves can be closely reproduced using the linear bias of \cite{mo}, in agreement with \cite{iliev2}.
% for a more detailed comparison between linear and nonlinear biases.

% is redshift dependence evident here? low redshift?

Our main results are shown in the bottom panel of Fig. \ref{biases}, which shows the redshift variation of the \ii{rms} 21cm fluctuations \re{fluct} for $\fnl\leq1$. The peak structure of these curves is the result of the competition between terms in Eq. \ref{fluct}, with $\sigma_p$ and $\overline{\delta T_b}$ decreasing with $z$ (see fig. 2 of \cite{iliev}), whlist $\beta$ increases as shown in the top panel. In this figure, we assume an observation bandwidth $\Delta\nu=1 $ MHz and beam angular diameter $\Delta\theta=$9 arcminutes. Also overlaid are two increasing curves corresponding to the noise \citep{furlanettoreview}
\be \delta T\sub{noise}\approx 20 \mbox{ mK } {10^4 \mbox{m}^2\over A\sub{tot}}\bkts{10^\pr \over \Delta\theta}^2\bkts{1+z\over 10}^{4.6}\bkts{{\mbox{MHz}\over\Delta\nu} {100 \mbox{hr} \over t\sub{int}} }^{1/2}\nn\ee
The noise thresholds assume total effective areas $A\sub{tot}=10^4$ m$^2$ (``LOFAR") and $10^5$ m$^2$ (``SKA"), with integration time $t\sub{int}=1000$ hours in both cases (see \cite{devos,dewdney} for detailed specifications). These curves show that even $\fnl$ as small as $0.1$ will boost the fluctuations from a few mKs to tens of mK. Hence, there are good prospects of detecting a small NG  via the 21cm emission from high-redshift minihalos with upcoming radio telescopes.

%%%%%%%%%%%%%%%%%%%%%%%%%%

%%%%%%%%%%%%%%%%%%%%%%%%%%
  
%For illustration, we also consider a futuristic imaging array with $A\sub{tot}=10^6$ m$^2$ and resolution $\Delta\theta=1$ arcminute. The overlaid curves are noise thresholds for $t\sub{int}=10^3$ and $10^4$ hrs. In both cases, we see that there are realistic prospects of detecting a small primordial NG ($\fnl=\mc{O}(1)$) via the enhancement in the 21cm-emission fluctuations from high-redshift minihalos.

Moreover, we have checked that the non-Gaussian effects on $\bkta{\delta T_b^2}$ cannot be easily reproduced by changing each of the fiducial cosmological parameters (within the observational limits). For instance, increasing $\sigma_8$ from 0.8 to 0.9 results in $\lesssim0.1$mK increase across the redshift range shown. A Fisher matrix analysis could be performed to shed light on parameter correlations, but we shall leave this for a future investigation.

\section{Discussion}\lab{later}
% CAVEATS

Here we consider a number of caveats for the results in Fig. \ref{biases}, mainly coming from the fact that MHs are relatively small objects whose dynamics are governed by nonlinear physics on small scales. 

\sss

\no\ii{i) Mass Function:} MHs are notoriously difficult to resolve in $N$-body simulations, requiring at least a $\sim20$ Mpc comoving box and $\gtrsim10^{10}$ Jeans-mass particles \citep{meiksin} (see \cite{iliev2,shapiro,ciardi} for previous attempts). The large simulation in \cite{iliev4}, in particular, was able to resolve down to $10^4M_\sun$ MHs and found their abundance to  lie between the Press-Schechter \citep{ps} and Sheth-Tormen \citep{st} predictions. We now consider how the 21cm emission from MHs is affected by the choice of mass function.

\begin{figure}
%\vskip -1 in
\centering
\includegraphics[width= 11cm, angle = -90]{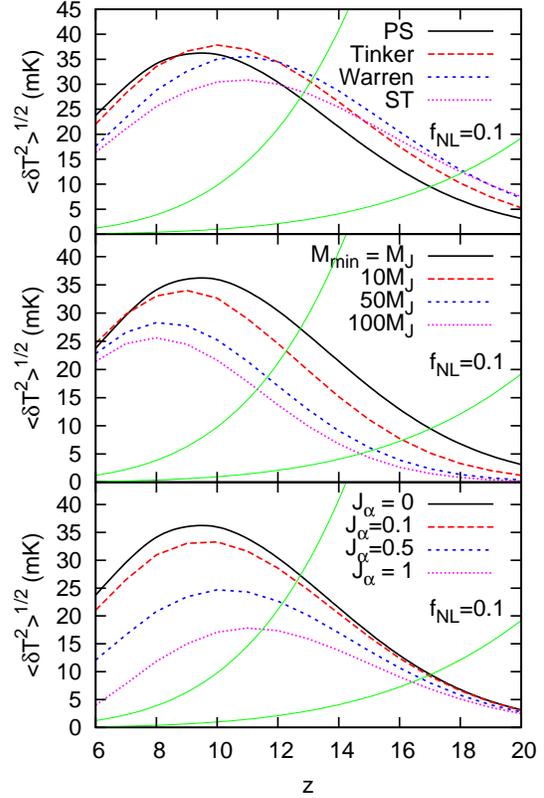}
\caption{Fluctuations in the 21cm brightness temperature, assuming the same telescope specifications as the bottom panel of Fig. \re{biases} with $\fnl=0.1$. The ``LOFAR" and ``SKA" noise curves in that figure are also reproduced here (note the linear scales). The fiducial model (black/solid) assumes the Press-Schechter mass function, MH mass threshold $M\sub{min}= M_J$ and radiation intensity (at Ly$\alpha$ frequency) $J_\alpha=0$. The panels show deviations from this model when the mass function (top), $M\sub{min}$ (middle) or $J_\alpha$ (bottom) is varied.}
\label{XXX}

\end{figure}

In the top panel of Fig. \ref{XXX}, we replot the 21cm fluctuations in the lower  panel of Fig. \ref{biases} for $\fnl=0.1$ using the above mass functions along with those of \cite{tinker} and \cite{warren} (note the linear scale here). The Tinker mass function is known to predict $n(z)$ lying between the Press-Schechter and the Sheth-Tormen predictions (\eg \cite{me}). The Warren mass function was found to be accurate for high-redshift objects down to $10^7h^{-1}M_\sun$ when compared with simulations \citep{lukic}. We see that the Press-Schechter and Tinker prescriptions gave similarly high amplitudes of $\bkta{\delta T_b^2}$ for $z\lesssim10$, whilst the Warren and Sheth-Tormen prescriptions prefer lower amplitudes. The trends are reversed for $z\sim20$. In any case, the uncertainty in the mass function does not appear to affect the detection prospects for LOFAR and SKA.
%which will most likely elude the first-generation arrays like LOFAR unless $\fnl>$ a few.

%%%%%%%%%%%%%%%%%%%%%%%%%%

%%%%%%%%%%%%%%%%%%%%%%%%%%

\sss

\no{\ii{ii) Uncertainty in} $M\sub{min}$}: On MH scales there are a number of so-called ``gastrophysical'' effects which may overwhelm the imprints of NG. For instance, \cite{tseliakhovich} showed that for $z\gtrsim10$, the velocity of dark matter relative to baryons is generally supersonic and 
 thus baryons can advect out of dark matter potential, leading to the possibility that $M\sub{min}>M_J$. This may be further exacerbated by feedback mechanisms such as photoevaporation of MHs \citep{iliev5} and shock heating in the IGM \citep{oh,furlanettoloeb}. 

The centre panel of Fig. \ref{XXX} shows the effects on $\bkta{\delta T_b^2}$ when  $M\sub{min}$ increases by a factor of 10, 50 and 100 (with $\fnl=0.1$). The result is a suppression in $\bkta{\delta T_b^2}$ across all redshifts (although gastrophysical suppressions are expected to be more significant at $z\gtrsim10$.) Nevertheless, we see that the fluctuation amplitudes are generally robust against changes in $M\sub{min}$.

\sss

\no\ii{iii) Ly$\alpha$ pumping}: The 21cm signals from MHs are unlikely to be completely immune to the Wouthuysen-Field mechanism, which redistributes the spin states and couples the spin temperature to that of radiation sources. This means that as the radiation intensity increases, $T_S\rightarrow T_K$ and the MH emission will be more and more suppressed.

Following \cite{chuzhoy}, we introduce the Ly$\alpha$ coupling of the form
\be y_\alpha =  1.3\times10^{-12} \bkts{J_\alpha T_*\over A_{10}T_K} {\exp\bkt{-0.3(1+z)^{1/2}T_K^{-2/3}} \over 1+0.4T_K^{-1}}, \ee
where $J_\alpha$ parametrizes the intensity of the radiation sources at the Ly$\alpha$  frequency (in units of $10^{-21}$ erg cm$^{-2}$ s$^{-1}$ Hz$^{-1}$ sr$^{-1}$). In the bottom panel of Fig. \ref{XXX}, we plot the effects of $J_\alpha=0.1, 0.5,$ and 1 on the 21cm fluctuations. Indeed we see a strong suppression of $\bkta{\delta T_b^2}$, with $J_\alpha\gtrsim1$ capable of effectively cancelling the boost from NG.

However, a more serious issue concerning Ly$\alpha$ pumping is that the MH signals will be completely submerged under a huge \ii{absorption} signal from the IGM, which contains much more mass than that in MHs \citep{ohmack, furlanettooh, yue}. In our example with $J_\alpha = 0.1$, we find an absorption amplitude of $|\bkta{\delta T_b^2}|\sim20$ mK, increasing to $100$ mK when $J_\alpha=1$. We therefore conclude that if MHs are indeed exposed to strong Ly$\alpha$ pumping, their 21cm emission will not be visible unless $\fnl\gg1$.

\section{Conclusions}

We have shown that a small amplitude of primordial non-Gaussianity may be detectable via the fluctuations in the 21cm emission of high-redshift MHs. Even with $\fnl\lesssim1$, we showed that the strong enhancement in the bias leads to a significant increase in the amplitude of the fluctuations, as seen in Fig. \ref{biases}. There are good prospects for such a detection by the next generation of large radio telescopes such as the SKA.

Our conclusions rely on a number of assumptions on the physics of MHs at high redshift. The analytic formalism used to calculate the bias assumes some extrapolations from cluster scales on which the theory has been well-tested. An improved calculation awaits a fuller understanding of nonlinear collapse in the presence of NG, perhaps with help from the resurgence of interests in excursion set theory. The results presented here are robust against the assumed mass function and minimum MH mass threshold. However, the fluctuations are sensitive to the presence of radiation background, since the Wouthuysen-Field effect is capable of overwhelming the MH signals with the IGM line-absorption signals.

Nevertheless, our conclusions still apply to MHs that are sufficiently isolated from UV sources with no strong feedback. This class of MHs provides a new probe for primordial non-Gaussianity which has so far been unexplored. For MHs that are subjected to strong background radiation, it may be possible for the MH signals to dominate if non-Gaussianity is much larger than the CMB bounds (perhaps with $\fnl\sim$100). However, this awaits high-resolution simulations to elucidate  high-redshift gastrophysics, cosmic reionization and the nonlinear bias given highly non-Gaussian initial conditions. Needless to say, this will be an extremely challenging task.

% mergers? reionization physics?

\mmm

We thank Patrick Valageas and, in particular, Ilian Iliev for illuminating discussions. We also thank the referee for comments that led to major improvements of the first version. SC is grateful for the support of Lincoln College, Oxford, where part of this work was completed.

\bibliographystyle{mn2e}
\bibliography{21cmbib}

\end{document}